# Particular and unique solutions of DGLAP evolution equations and light sea and valence quark structure functions at low-x

R.Rajkhowa[1] and J.K. Sarma[2]
[1]Physics Department, T. H. B. College, Jamugurihat, Sonitpur, Assam, India
[2]Physics Department, Tezpur University Napaam, Tezpur-784 028, Assam, India
[1]E-mail:rasna@tezu.ernet.in, [2]E-mail:jks@tezu.ernet.in

**Abstract**

We present particular and unique solutions of Dokshitzer-Gribov-Lipatov- Altarelli-Parisi (DGLAP) evolution equations for light sea and valence quark structure functions in leading order (LO). We obtain *t* evolutions of sea and valence quark structure functions and *x*- evolutions of light sea quark structure functions from DGLAP evolution equations. The results are compared with a recent global parameterization.

## 1. Introduction

In recent papers [1-3], particular solutions of the Dokshitzer-Gribov-Lipatov-Altarelli-Parisi (DGLAP) [4-7] evolution equations for *t* and *x*-evolutions of singlet and non-singlet structure functions in leading order (LO) and next-to-leading order (NLO) at low-*x* have been reported. In this paper we obtain particular and unique solutions of the DGLAP evolution equations for sea and valence quark structure functions in LO to obtain *t* and *x*-evolutions of those structure functions. These LO results are compared with a recent global parameterization [8]. Here section 1, section 2, and section 3 will give the introduction, the necessary theory and the results and discussion respectively.

## 2. Theory

The DGLAP evolution equations for sea and valence quark structure functions in the standard forms are [9]

$$\frac{\partial F_2^S\left(x,Q^2\right)}{\partial \ln Q^2}-\frac{2\alpha_S(Q^2)}{3\pi}\left[\int_x^1\frac{dw}{1-w}\left\{\left(1+w^2\right)F_2^S\left(x/w,Q^2\right)-2F_2^S\left(x,Q^2\right)\right\}\right]+\frac{2\alpha_S(Q^2)}{3\pi}\left[1+\frac{4\ln(1-x)}{3}\right]F_2^s(x,Q^2) \quad (1)$$

$$\frac{3}{8}\left[w^2+(1+w)^2\right]G(x/w,Q^2)=0$$

and

$$\frac{\partial F_2^v(x,Q^2)}{\partial \ln Q^2}-\frac{2\alpha_s(Q^2)}{3\pi}\left[\int_x^1\frac{dw}{1-w}\left\{\left(1+w^2\right)F_2^v\left(x/w,Q^2\right)-2F_2^v\left(x,Q^2\right)\right\}\right]+\frac{2\alpha_s(Q^2)}{3\pi}\left[1+\frac{4\ln(1-x)}{3}\right]F_2^v(x,Q^2)=0, \quad (2)$$

where $F_2^s(x, Q^2) = xu_s$ or $xd_s$ or $xs_s$, $F_2^v(x, Q^2) = xu_v$ or $xd_v$ and $\alpha_s(Q^2)=\frac{33-2N_f}{12\pi}\ln\left(\frac{Q^2}{\Lambda^2}\right)$. Taking $t = \ln(Q^2/\Lambda^2)$, $A_f = 4/(33-2N_f)$, $N_f$ being the number of flavours and $\Lambda$ is the QCD cut off parameter.

Eqs (1) and (2) become,

$$\frac{\partial F_2^s(x,t)}{\partial t}-\frac{A_f}{t}\left[\{3+4\ln(1-x)\}F_2^s(x,t)+I_1^s(x,t)+I_2^s(x,t)\right]=0 \quad (3)$$

and

$$\frac{\partial F_2^v(x,t)}{\partial t}-\frac{A_f}{t}\left[\{3+4\ln(1-x)\}F_2^v(x,t)+I^v(x,t)\right]=0, \quad (4)$$

where

$$I_1^s(x,t)=2\int_x^1\frac{dw}{1-w}\left\{\left(1+w^2\right)F_2^s(x/w,t)-2F_2^s(x,t)\right\}, \quad (5)$$

$$I_2^s(x,t)=\frac{3}{4}\int_x^1\left\{w^2+(1-w)^2\right\}G(x/w,t)dw \quad (6)$$

and

$$I^v(x,t)=2\int_x^1\frac{dw}{1-w}\left\{\left(1+w^2\right)F_2^v(x/w,t)-2F_2^v(x,t)\right\}. \quad (7)$$

Using Taylor expansion method [10] and neglecting higher order terms of $x$ as discussed in our earlier works [1-3, 11-12], $G(x/w,t)$ can be approximated for low-$x$ as

$$G(x/w,t)\cong G(x,t)+x\sum_{k=1}^{\infty}u^k\frac{\partial G(x,t)}{\partial x}. \quad (8)$$

where, $u=1-w$ and $\frac{x}{1-u}=x\sum_{l=0}^{\infty}u^l$ .

Similarly, $F_2^s(x/w, t)$ and $F_2^v(x/w, t)$ can be approximated for small-$x$ as

$$F_2^s(x/w,t)\cong F_2^s(x,t)+x\sum_{k=1}^{\infty}u^k\frac{\partial F_2^s(x,t)}{\partial x} \quad (9)$$

and

$$F_2^{v}\left(x/w,t\right) \cong F_2^{v}\left(x,t\right) + x\sum_{k=1}^{\infty}u^{k}\frac{\partial F_2^{v}\left(x,t\right)}{\partial x}. \tag{10}$$

Using equations (8), (9) and (10) in equations (5), (6) and (7) and performing *u*-integrations we get

$$I_1^{S} = -[(1-x)(x+3)]F_2^{S}(x,t) + \left[2x\ln(1/x) + x\left(1-x^2\right)\right]\frac{\partial F_2^{S}(x,t)}{\partial x}, \tag{11}$$

$$I_2^{S} = \left[\frac{1}{4}(1-x)\left(2-x+2x^2\right)G(x,t) + \left\{-\frac{1}{4}x(1-x)\left(5-4x+2x^2\right) + \frac{3}{4}x\ln(1/x)\right\}\frac{\partial G(x,t)}{\partial x}\right] \tag{12}$$

and

$$I^{V} = -[(1-x)(x+3)]F_2^{V}(x,t) + \left[2x\ln(1/x) + x\left(1-x^2\right)\right]\frac{\partial F_2^{V}(x,t)}{\partial x}.. \tag{13}$$

Now using equations (11) and (12) in equation (1) we have,

$$\frac{\partial F_2^{S}(x,t)}{\partial t} - \frac{A_f}{t}\left[A(x)F_2^{S}(x,t) + B(x)\frac{\partial F_2^{S}(x,t)}{\partial x} + C(x)G(x,t) + D(x)\frac{\partial G(x,t)}{\partial x}\right] = 0. \tag{14}$$

Let us assume for simplicity,

$G(x, t) = K(x)\, F_2^{s}\,(x, t),$ (15)

where $K(x)$ is a function of $x$. Now equation (14) gives

$$\frac{\partial F_2^{S}(x,t)}{\partial t} - \frac{A_f}{t}[L(x)F_2^{S}(x,t) + M(x)\frac{\partial F_2^{S}(x,t)}{\partial x}] = 0, \tag{16}$$

where

$A(x)= 3+ 4 \ln(1-x)-(1-x)(3+ x)$, $B(x) = x\,(1-x^2) + 2x \ln(1/ x)$, $C(x) = 1/ 4\, (1-x)( 2-x+ 2x^2)$,

$D(x) = x\,[-1/ 4(1-x)(5-4x+ 2x^2)+(3/ 4) \ln(1/ x)$, $L(x) = A(x) + K(x)C(x) + D(x)\frac{\partial K(x)}{\partial x}$

and $M(x) = B(x)+ K(x)D(x)$.

Secondly, using equation (13) in equation (2) we have

$$\frac{\partial F_2^{V}(x,t)}{\partial t} - \frac{A_f}{t}\left[P(x)F_2^{V}(x,t) + Q(x)\frac{\partial F_2^{V}(x,t)}{\partial x}\right] = 0, \tag{17}$$

where $P(x) = 3+ 4 \ln(1-x)-(1-x)(x+ 3)$ and $Q(x) = x(1-x^2)-2x \ln x$. The general solutions of equations (16) is [13-14] $F\,(U, V) = 0$, where $F$ is an arbitrary function and $U(x, t, F_2) = C_1$ and $V(x, t, F_2) = C_2$ form a solution of equation

$$\frac{dx}{A_f M(x)} = \frac{dt}{-t} = \frac{dF_2^S(x,t)}{-A_f L(x) F_2^s(x,t)}. \tag{18}$$

Solving equation (20) we obtain, $U\left(x,t,F_2^s\right) = t\exp\left[\frac{1}{A_f}\int\frac{1}{M(x)}dx\right]$ and $V\left(x,t,F_2^s\right) = F_2^S(x,t)\exp\left[\int\frac{L(x)}{M(x)}dx\right]$.

If $U$ and $V$ are two independent solutions of equation (18) and if $\alpha$ and $\beta$ are arbitrary constants, then $V = \alpha U + \beta$ may be taken as a complete solution of equation (16). We take this form as this is the simplest form of a complete solution which contains both the arbitrary constants $\alpha$ and $\beta$. Now the complete solution [13-14]

$$F_2^s(x,t)\exp\left[\int\frac{L(x)}{M(x)}dx\right] = \alpha t\exp\left[\frac{1}{A_f}\int\frac{1}{M(x)}dx\right] + \beta \tag{19}$$

is a two-parameter family of surfaces. The one parameter family determined by taking $\beta = \alpha^2$ has equation

$$F_2^s(x,t)\exp\left[\int\frac{L(x)}{M(x)}dx\right] = \alpha t\exp\left[\frac{1}{A_f}\int\frac{1}{M(x)}dx\right] + \alpha^2. \tag{20}$$

Differentiating equation (20) with respect to $\alpha$, we get $\alpha = -\frac{1}{2}t\exp\left[\frac{1}{A_f}\int\frac{1}{M(x)}dx\right]$. Putting the value of $\alpha$ in equation (20), we obtain the envelope

$$F_2^S(x,t) = -\frac{1}{4}t^2\exp\left[\int\left(\frac{2}{A_f M(x)} - \frac{L(x)}{M(x)}\right)dx\right], \tag{21}$$

which is merely a particular solution of the general solution. Now, defining

$F_2^S(x,t) = -\frac{1}{4}t_0^2\exp\left[\int\left(\frac{2}{A_f M(x)} - \frac{L(x)}{M(x)}\right)dx\right]$, at $t = t_0$, where $t_0 = \ln(Q_0^2/\Lambda^2)$ at any lower value $Q = Q_0$, we get from equation (21)

$$F_2^S(x,t) = F_2^S(x_0,t)\left(\frac{t}{t_0}\right)^2, \tag{22}$$

which gives the $t$-evolution of light sea quark structure function $F_2^s(x, t)$.

Proceeding exactly in the same way, and defining $F_2^v(x,t_0) = -\frac{1}{4}t_0^2\exp\left[\int\left(\frac{2}{A_f Q(x)} - \frac{P(x)}{Q(x)}\right)dx\right]$, we get for valence quark structure function

$$F_2^v(x,t) = F_2^v(x,t_0)\left(\frac{t}{t_0}\right)^2, \tag{23}$$

which gives the $t$-evolution of valence quark structure function $F_2^v(x, t)$.

Again defining, $F_2^S(x_0,t) = -\frac{1}{4}t^2 \exp\left[\int\left(\frac{2}{A_f M(x)} - \frac{L(x)}{M(x)}\right)dx\right]_{x=x_0}$, we obtain from equation (21)

$$F_2^S(x,t) = F_2^S(x_0,t)\exp\left[\int_{x_0}^{x}\left(\frac{2}{A_f M(x)} - \frac{L(x)}{M(x)}\right)dx\right], \tag{24}$$

which gives the $x$-evolution of light sea quark function $F_2^s(x, t)$. Similarly defining,

$F_2^v(x_0,t) = -\frac{1}{4}t^2 \exp\left[\int\left(\frac{2}{A_f Q(x)} - \frac{P(x)}{Q(x)}\right)dx\right]_{x=x_0}$, we get

$$F_2^v(x,t) = F_2^v(x_0,t)\exp\left[\int_{x_0}^{x}\left(\frac{2}{A_f Q(x)} - \frac{P(x)}{Q(x)}\right)\right]dx, \tag{25}$$

which gives the $x$-evolution of valence quark structure function $F_2^v(x, t)$.

For the complete solution of equation (16), we take $\beta = \alpha^2$ in equation (19). If we take $\beta = \alpha$ in equation (19) and differentiating with respect to $\alpha$ as before, we get

$$0 = t\exp\left[\frac{1}{A_f}\int\frac{1}{M(x)}dx\right] + 1$$

from which we can not determine the value of $\alpha$. But if we take $\beta = \alpha^3$ in equation (19) and differentiating with respect to $\alpha$, we get $\alpha = \sqrt{-\frac{1}{3}t\exp\left[\frac{1}{A_f}\int\frac{1}{M(x)}dx\right]}$ which is imaginary. Putting this value of $\alpha$ in equation (19) we get ultimately $F_2^S(x,t) = t_S^{\frac{3}{2}}\left\{\left(-\frac{1}{3}\right)^{\frac{1}{2}} + \left(-\frac{1}{3}\right)^{\frac{3}{2}}\right\}\exp\left[\int\left(\frac{\frac{3}{2}}{A_f M(x)} - \frac{L(x)}{M(x)}\right)dx\right]$.

Now, defining

$F_2^S(x,t_0) = t_0^{\frac{3}{2}}\left\{\left(-\frac{1}{3}\right)^{\frac{1}{2}} + \left(-\frac{1}{3}\right)^{\frac{3}{2}}\right\}\exp\left[\int\left(\frac{\frac{3}{2}}{A_f M(x)} - \frac{L(x)}{M(x)}\right)dx\right]$ we get,

$F_2^S(x,t) = F_2^S(x,t_0)\left(\frac{t}{t_0}\right)^{\frac{3}{2}}$, which gives the $t$-evolution of light sea quark structure function $F_2^s(x, t)$.

Proceeding exactly in the same way we get for valence quark structure function also

$F_2^V(x,t) = F_2^V(x,t_0)\left(\frac{t}{t_0}\right)^{\frac{3}{2}}$, which gives the $t$-evolution of valence quark structure function $F_2^v(x, t)$.

Proceeding in the same way we get $x$- evolutions of light sea and valence quark structure functions as

$$F_2^S(x,t) = F_2^S(x_0,t)\exp\left[\int_{x_0}^{x}\left(\frac{\frac{3}{2}}{A_f M(x)} - \frac{L(x)}{M(x)}\right)dx\right] \quad \text{and} \quad F_2^V(x,t) = F_2^V(x_0,t)\exp\left[\int_{x_0}^{x}\left(\frac{\frac{3}{2}}{A_f Q(x)} - \frac{P(x)}{Q(x)}\right)dx\right]$$

respectively.

Proceeding exactly in the same way we can show that if we take $\beta = \alpha^4$ we get

$$F_2^S(x,t) = F_2^S(x,t_0)\left(\frac{t}{t_0}\right)^{\frac{4}{3}}, \quad F_2^S(x,t) = F_2^S(x_0,t)\exp\left[\int_{x_0}^{x}\left(\frac{\frac{4}{3}}{A_f M(x)} - \frac{L(x)}{M(x)}\right)dx\right] \text{ and}$$

$$F_2^V(x,t) = F_2^V(x,t_0)\left(\frac{t}{t_0}\right)^{\frac{4}{3}}, \quad F_2^V(x,t) = F_2^V(x_0,t)\exp\left[\int_{x_0}^{x}\left(\frac{\frac{4}{3}}{A_f Q(x)} - \frac{P(x)}{Q(x)}\right)dx\right]dx,$$

and so on. So in general, if we take $\beta = \alpha^y$, we get

$$F^{S,V}(x,t) = F^{S,V}(x,t_0)\left(\frac{t}{t_0}\right)^{\frac{y}{y-1}}, \quad F^S(x,t) = F^S(x_0,t)\exp\left[\int_{x_0}^{x}\left(\frac{\frac{y}{y-1}}{A_f M(x)} - \frac{L(x)}{M(x)}\right)dx\right] \text{ and}$$

$F^V(x,t) = F^V(x_0,t)\exp\left[\int_{x_0}^{x}\left(\frac{\frac{y}{y-1}}{A_f Q(x)} - \frac{P(x)}{Q(x)}\right)dx\right]$, which our $t$ and $x$-evolutions respectively of light sea and valence quark structure functions for $\beta = \alpha^y$. We observe if $y \to \infty$ (very large), $y/(y\text{-}1) \to 1$.

Thus we observe that if we take $\beta = \alpha$ in equation (19) we can not obtain the value of $\alpha$ and also the required solution. But if we take $\beta = \alpha^2, \alpha^3, \alpha^4, \alpha^5$….. and so on, we see that the powers of $(t/t_0)$ in $t$-evolutions and the numerators of the first term inside the integral for $x$- evolutions of valence and light sea quark structure functions are 2, 3/2, 4/3, 5/4….and so on respectively as discussed above. Thus we see that if in the relation $\beta = \alpha^y$, $y$ varies between 2 to a maximum value, the powers of $(t/t_0)$ varies between 2 to 1, and the numerator of the first term in the integral sign varies between 2 to 1. Then it is understood that the solutions of equations (16) and (17) obtained by this methodology are not unique and so the *t*-evolutions and *x*-evolution of valence and light sea quark structure function obtained by this methodology are not unique. Thus by this methodology, instead of having a single solution we arrive a band of solutions, of course the range for these solutions is reasonably narrow.

Again due to conservation of the electromagnetic current, $F_2$ must vanish as $Q^2$ goes to zero [15, 16]. Also $R \to 0$ in this limit. Here $R$ indicates ratio of longitudinal and transverse cross-sections of virtual photon in DIS process. This implies that scaling should not be a valid concept in the region of very low $Q^2$. The exchanged photon is then almost real and the close similarity of real photonic and hadronic interactions justifies the use of the Vector Meson Dominance (VMD) concept [17-18] for the description of $F_2$. In the language of perturbation theory this concept is equivalent to a statement that a physical photon spends part of its time as a "bare", point-like photon and part as a virtual hadron (s) [16]. The power and beauty of explaining scaling violations with field theoretic methods (i.e., radiative corrections in QCD) remains, however, unchallenged in as much as they provide us with a framework for the whole *x*-region with essentially only one free parameter $\Lambda$ [19]. For $Q^2$ values much larger than $\Lambda^2$, the effective coupling is small and a perturbative description in terms of quarks and gluons interacting weakly makes sense. For $Q^2$ of order $\Lambda^2$, the effective coupling is infinite and we cannot make such a picture, since quarks and gluons will arrange themselves into strongly bound clusters, namely, hadrons [15] and so the perturbation series breaks down at small-$Q^2$ [15, 20]. Thus, it can be thought of $\Lambda$ as marking the boundary between a world of quasi-free quarks and gluons, and the world of pions, protons, and so on. The value of $\Lambda$ is not predicted by the theory; it is a free parameter to be determined from experiment. It should expect that it is of the order of a typical hadronic mass [15]. Since the value of $\Lambda$ is so small we assume at $Q = \Lambda$, $F_2^S(x, t) = 0$ due to conservation of the electromagnetic current [15-16]. This dynamical prediction agrees with most ad hoc parameterizations and with the data [19]. Using this boundary condition in equation (19) we get $\beta = 0$ and

$$F_2^S(x,t) = \alpha t \exp\left[\int\left(\frac{1}{A_f M(x)} - \frac{L(x)}{M(x)}\right)dx\right]. \tag{26}$$

Now, defining $F_2^S(x,t) = \alpha t_0 \exp\left[\int\left(\frac{1}{A_f M(x)} - \frac{L(x)}{M(x)}\right)dx\right]$, at $t = t_0$, where, $t_0 = \ln(Q_0^2/\Lambda^2)$ at any lower value $Q = Q_0$, we get from eq (26)

$$F_2^S(x,t) = F_2^S(x, t)\left(\frac{t}{t_0}\right), \tag{27}$$

which gives the $t$-evolutions of light sea quark structure function in LO. Again defining,

$F_2^S(x_0,t)=\alpha t \exp\left[\int\left(\frac{1}{A_f M(x)}-\frac{L(x)}{M(x)}\right)dx\right]_{x=x_0}$, we obtain from eq (26)

$$F_2^S(x,t)=F_2^S(x_0,t)\exp\left[\int_{x_0}^{x}\left(\frac{1}{A_f M(x)}-\frac{L(x)}{M(x)}\right)dx\right], \tag{28}$$

which gives the $x$-evolutions of light sea quark structure functions in LO. Similarly we get for valence quark

$$F_2^V(x,t)=F_2^V(x,t_0)\left(\frac{t}{t_0}\right) \tag{29}$$

and

$$F_2^V(x,t)=F_2^V(x_0,t)\exp\left[\int_{x_0}^{x}\left(\frac{1}{A_f Q(x)}-\frac{P(x)}{Q(x)}\right)\right]dx. \tag{30}$$

We observed that unique solutions (equations (27), (28), (29) and (30)) of DGLAP evolution equations for valence and light sea quark structure functions are same with particular solutions for $y$ maximum in $\beta = \alpha^y$ relation in LO.

## 3. Results and discussion

In the present paper, we present our result of $t$-evolution of valence and light sea quark structure functions qualitatively and compare result of $x$-evolution with a recent global parameterization [8]. These parameterizations include data from H1, ZEUS, DO, CDF data. Though we present our results of $t$-evolution with $y$ = 2 and $y$ = maximum in $\beta = \alpha^y$ relation our result with $y$ = maximum is equivalent to that of unique solution and results of $x$-evolution for $y$ = 2 and $y$ = maximum in $\beta = \alpha^y$ relation have not any significant difference.

In fig.1 (a-c) we present our results of $t$-evolutions of light sea and valence quark structure functions qualitatively for the representative values of $x$ given in the figures for $y$ = 2 (solid lines) and $y$ maximum (dashed lines) in $\beta = \alpha^y$ relation. We have taken arbitrary inputs from recent global parameterizations MRST2001 [8] at $Q_0^2$= 1 GeV$^2$. It is clear from figures that $t$-evolutions of valence and light sea quark structure functions depend upon input $F^s(x, t_0)$ and $F^v(x, t_0)$ values. Unique solutions of $t$-evolution for light sea and valence quark structure functions are same with particular solutions for $y$ maximum in $\beta = \alpha^y$ relation in LO.

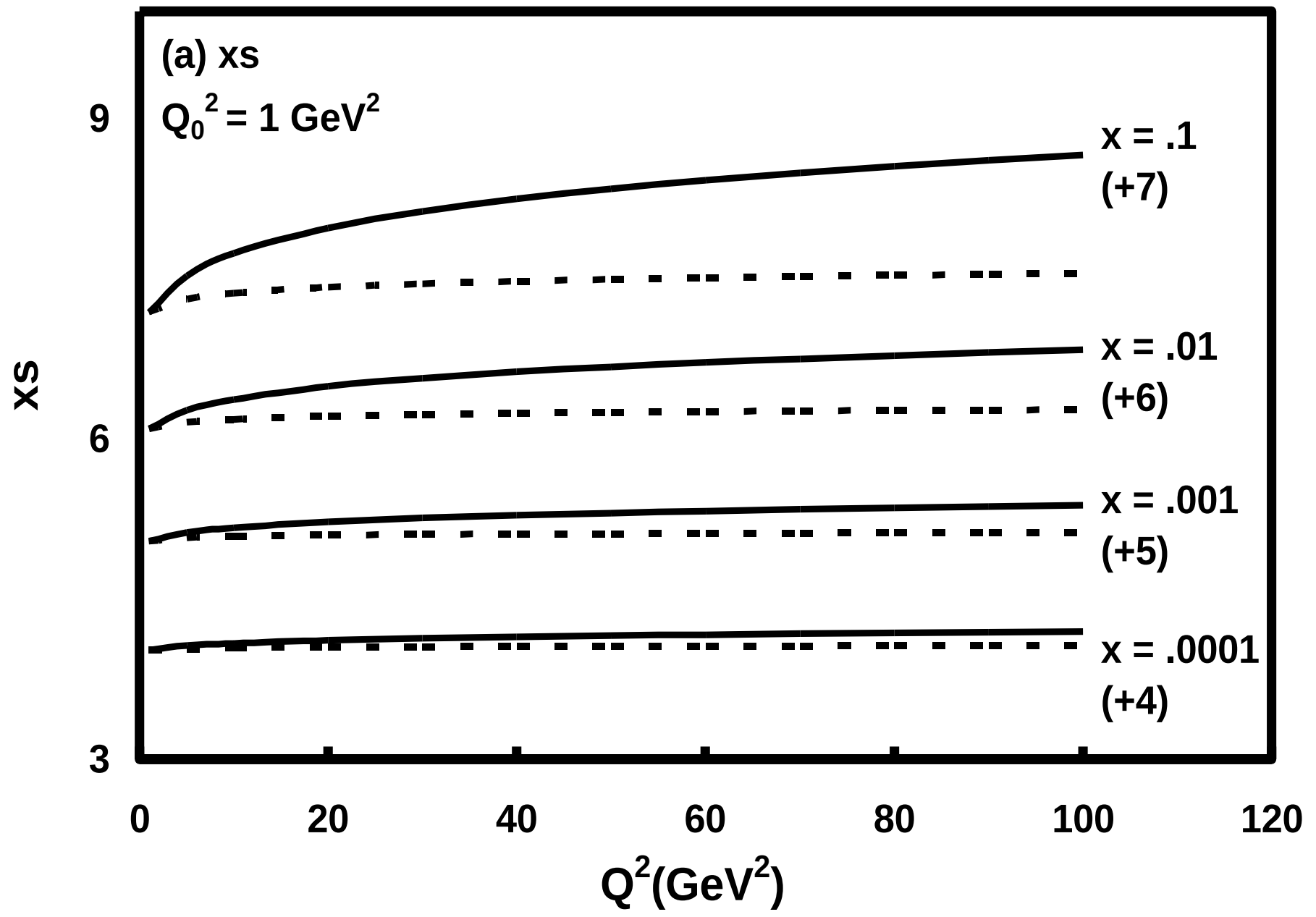

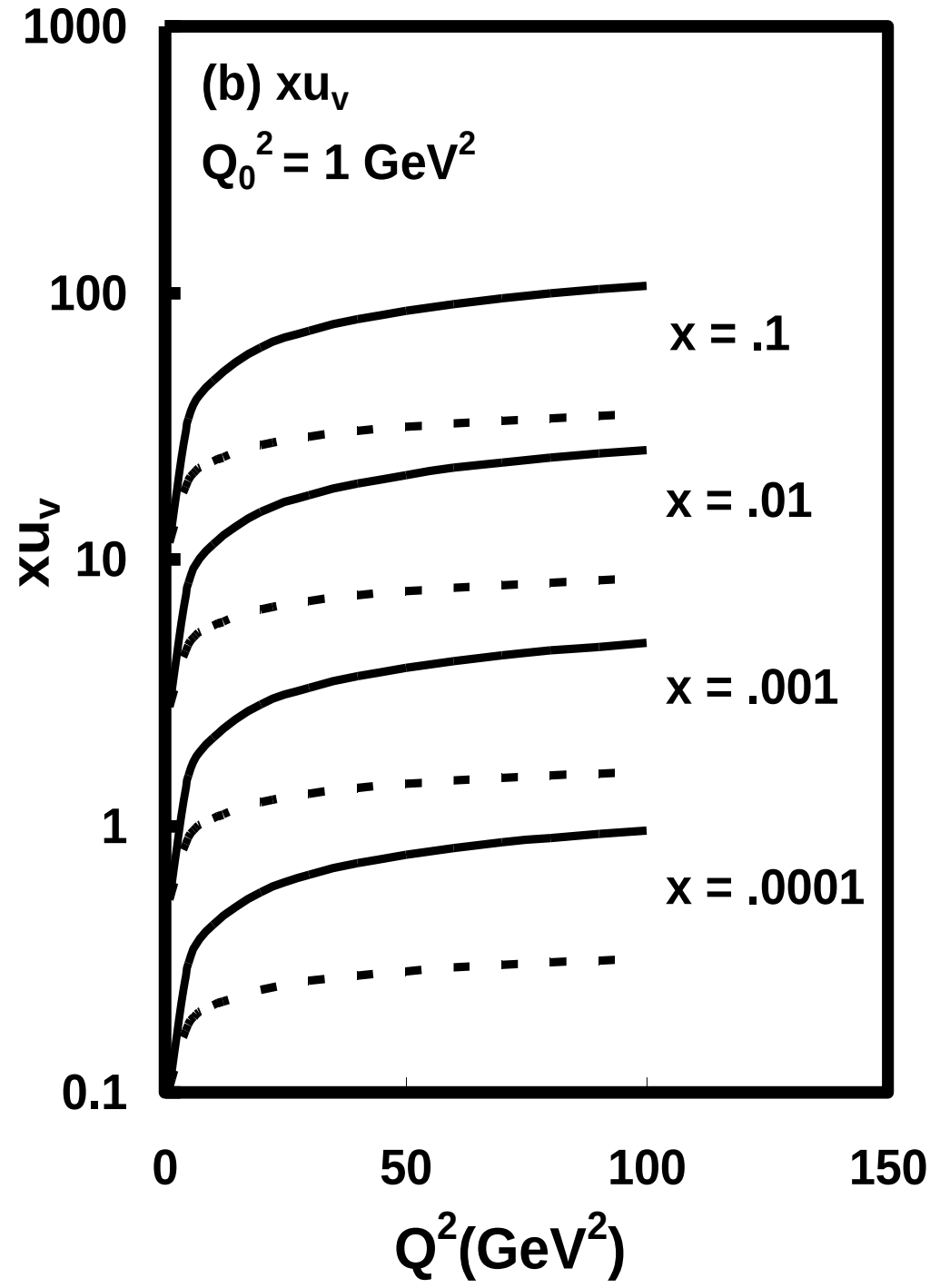

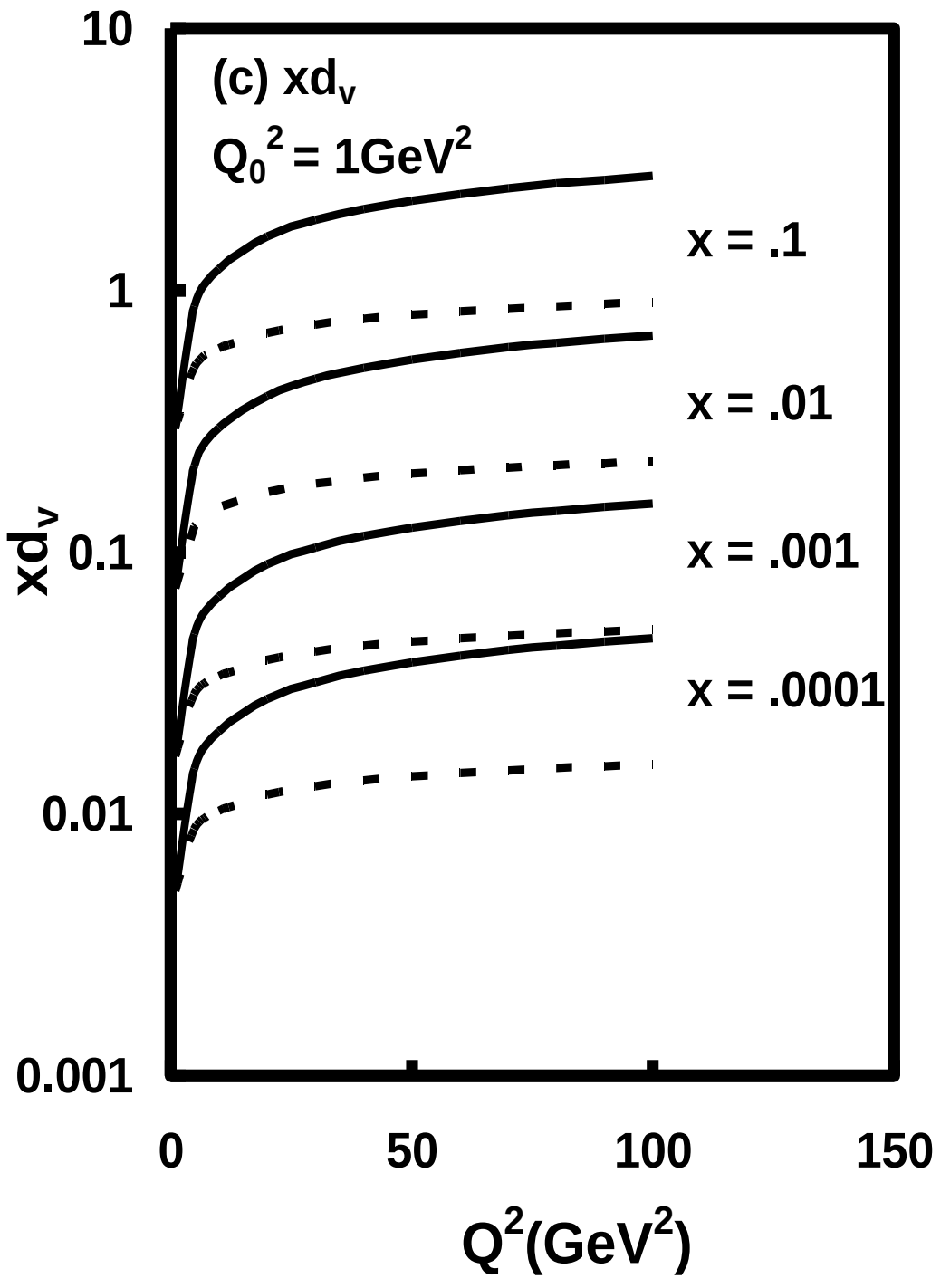

**Fig. 1(a-c):** Results of *t*-evolutions of light sea and valence quark structure functions qualitatively for the representative values of $x$ given in the figures for $y$ = 2 (solid lines) and $y$ maximum (dashed lines) in $\beta = \alpha^y$ relation. We have taken arbitrary inputs from recent global parameterizations MRST2001 [8] at $Q_0^2$= 1 GeV$^2$. For convenience, value of each data point is increased by adding 4, 5, 6, 7 for $x$ = 0.0001, 0.001, 0.01, 0.1 respectively.

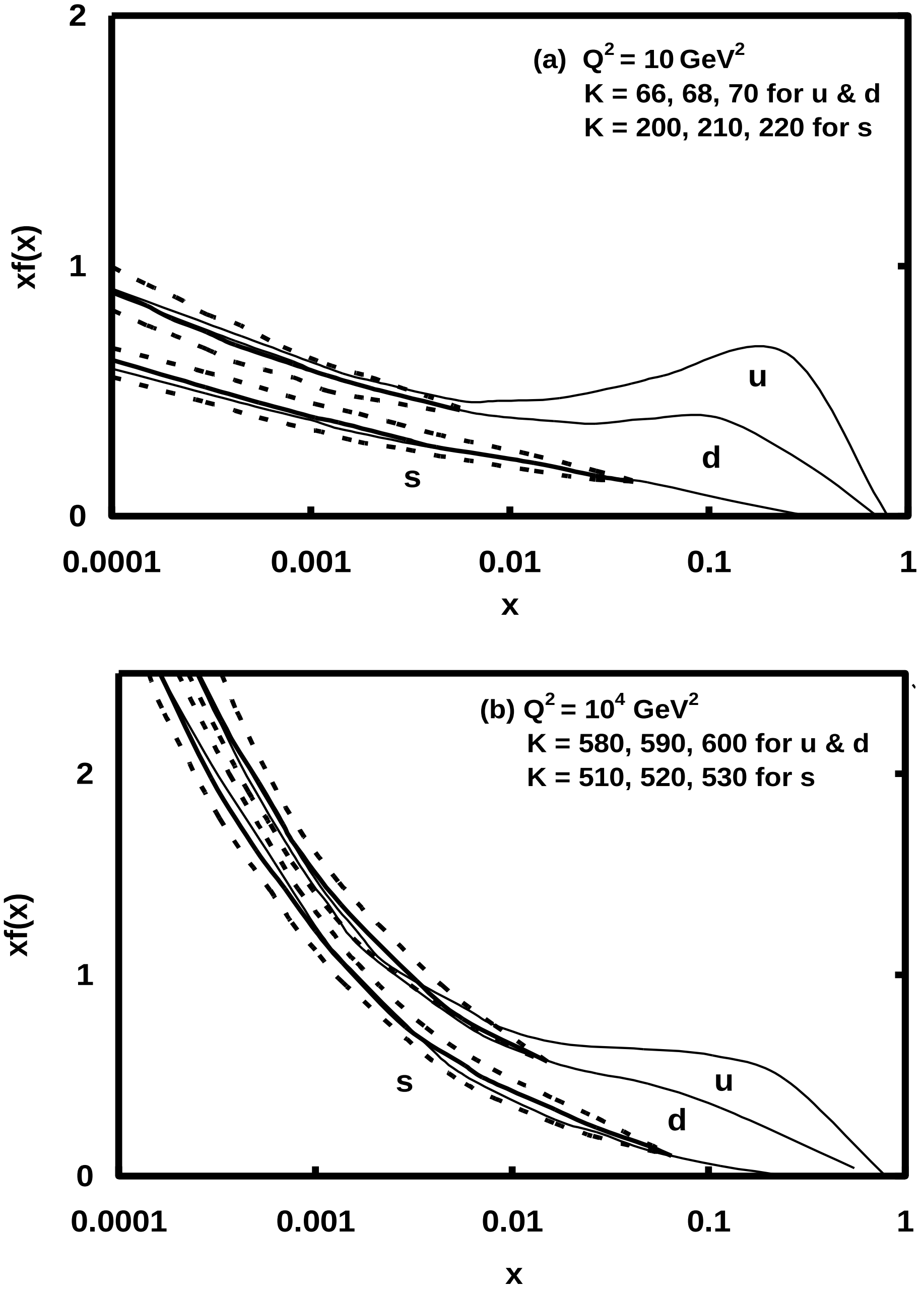

**Fig. 2(a-b):** Results of $x$-distribution of light sea quark structure functions for $K(x)$ = constant for representative values of $Q^2$ given in each figure, and compare them with recent global parameterizations (thin solid lines) [8] in the relation $\beta=\alpha^y$ for $y$ minimum (thick solid lines). In the same figures we present the sensitivity of our results (dashed lines) for different constant values of $K(x)$.

For a quantitative analysis of $x$-distributions of light sea quark structure functions, we calculate the integrals that occurred in equation (24) for $N_f = 4$. In figure 2 (a-b), we present our results of $x$-distribution of light sea quark structure functions for $K(x)$ = constant for representative values of $Q^2$ = 10 GeV$^2$ (figure 2(a)) and $Q^2$ = 10 GeV$^2$ (figure 2(b)) and compare them with recent global parameterizations (thin solid lines) [8] in the relation $\beta=\alpha^y$ for $y$ =2 (thick solid lines). Since our theory is in small-$x$ region and does not explain the peak portion for $u$ & $d$, so in each the data point for $x$- value just below 0.1 for $s$ and 0.01 for u & d has been taken as input to test the evolution equation (24). We observed that agreement of the results (thick solid line) with parameterization is found to be good at $K(x)$ = 60, 590 for $u$ & $d$ and $K(x)$ = 210, 520 for $s$ in figure 2(a) and figure 2(b) respectively. In the same figures we present the sensitivity of our results (dashed lines) for different constant values of $K(x)$. We observe that if value of $K(x)$ is increased or decreased, the curve goes upward or downward direction respectively. But the nature of the curve is similar.

In figures 3 (a-b) and 4(a-b) we present our results of $x$-distribution of light sea quark structure functions for $K(x) = ax^b$, where '$a$' and '$b$' are constants for representative values of $Q^2$ = 10 GeV$^2$ (figure 3(a-b)) and $Q^2$ = 10 GeV$^2$ (figure 4(a-b)) and compare them with recent global parameterizations (thin solid lines) [10] in the relation $\beta=\alpha^y$ for $y$ = 2 (thick solid lines). Since our theory is in small-$x$ region and does not explain the peak portion for $u$ & $d$, so in each the data point for $x$-value just below 0.1 for s and 0.01 for $u$ & $d$ has been taken as input to test the evolution equation (24). We observed that agreement of the results (thick solid line) with parameterization is found to be good at $a$ = 135 & $b$ = 0.33 for $u$ & $d$ and $a$ = 130 & $b$ = 0.35 for $s$ at $Q^2$ = 10 GeV$^2$ in figure 3(a-b) and $a$ = 211 & $b$ = 0.25 for $u$ & $d$ and $a$ = 260 & $b$ = 0.29 for s at $Q^2$ = $10^4$ GeV$^2$ in figure 4(a-b). In the same figures we present the sensitivity of our results (dashed lines) for different values of '$a$' and '$b$'. Here we take $b$ = 0.33, 0.35 in figure 3(a) and $b$ = 0.25, 0.29 in figure 4(a). We observe that if value of '$a$' is increased or decreased, the curve goes upward or downward direction. But the nature of the curve is similar.

In figure 3(b) and figure 4 (b), we present the sensitivity of our results (dashed lines) for different values of '$b$' at fixed value of '$a$'. Here we take $a$ = 135, 130 in figure 3(b) and $a$ = 211, 260 in figure 4(b). We observe that at $b$ = 0.33 & 0.35, agreement of the results (thick solid lines) with parameterizations data is found to be good in figure 3(b) and at $b$ = 0.25 & 0.29, agreement of the results (thick solid lines) with parameterizations data is found to be excellent in figure 4(b). If value of '$b$' is increased or decreased the curve goes downward or upward direction. But the nature of the curve is similar.

In figures 5(a-b) and 6(a-b) we present our results of $x$-distribution of light sea quark structure functions for $K(x) = ce^{-dx}$, where '$c$' and'$d$' are constants for representative values of $Q^2$ = 10 GeV$^2$ (figure 5(a-b)) and $Q^2$ = 10 GeV$^2$ (figure 6(a-b)) and compare them with recent global parameterizations (thin solid lines) [8] in the relation $\beta=\alpha^y$ for $y$ = 2 (thick solid lines). Since our theory is in small-$x$ region and does not explain the peak portion for $u$ & $d$, so in each the data point for $x$-value just below 0.1 for $s$ and

0.01 for *u* & *d* has been taken as input to test the evolution equation (24). We observed that agreement of the results (thick solid line) with parameterization is found to be good at *c* = 47.8 & *d* = -1 for *u* ,*d* and *c* =32.5 & *d*=-20 for *s* at $Q^2$ =10 $GeV^2$ in figs.5(a-b) and *c*=465 & *d*=-.4 for *u* ,*d* and *c*=385 & *d* = - 25 for $s$ at $Q^2 = 10^4$ GeV$^2$ in figs. 6(a-b). In the same figures we present the sensitivity of our results (by dashed lines) for different values of '$c$' and'$d$'. Here we take $d$ = -1, -20 in figure 5(a) and $d$ = -.4, -25 in figure 6(a). We observe that if value of '*c*' is increased or decreased, the curve goes upward or downward direction. But the nature of the curve is similar.

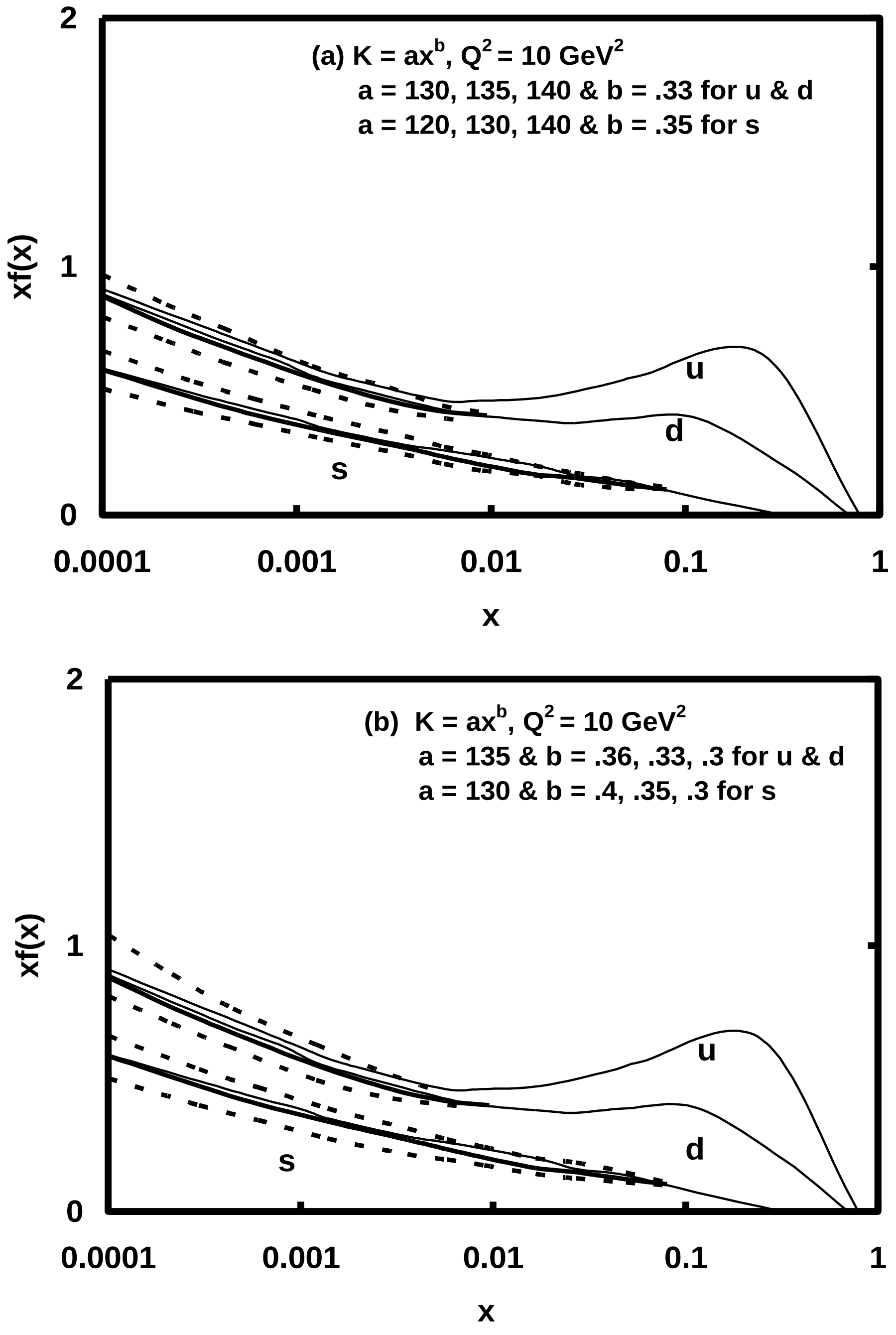

**Fig. 3(a-b):** Results of *x*-distribution of light sea quark structure functions for $K(x) = ax^b$, where '*a*' and '*b*' are constants for $Q^2$ = 10 $GeV^2$ and compare them with recent global parameterizations (thin solid lines) [8] in the relation $\beta=\alpha^y$ for *y* minimum (thick solid lines). In the same figures we present the sensitivity of our results (dashed lines) for different values of '*a*' and '*b*'.

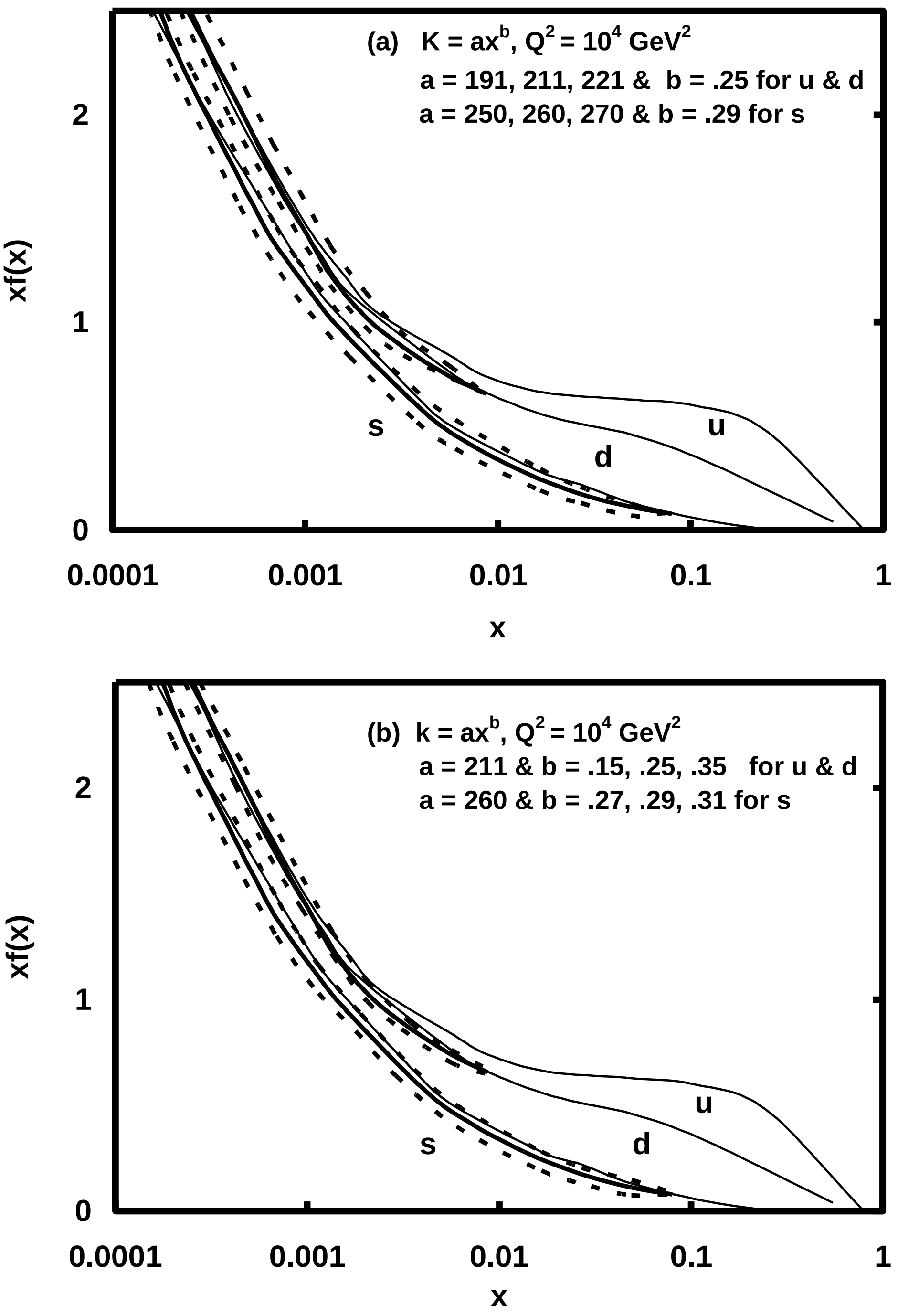

**Fig. 4(a-b):** Results of $x$-distribution of light sea quark structure functions for $K(x) = ax^b$, where '$a$' and '$b$' are constants for $Q^2 = 10^4$ GeV$^2$ and compare them with recent global parameterizations (thin solid lines) [10] in the relation $\beta = \alpha^y$ for $y$ minimum (thick solid lines). In the same figures we present the sensitivity of our results (dashed lines) for different values of '$a$' and '$b$'.

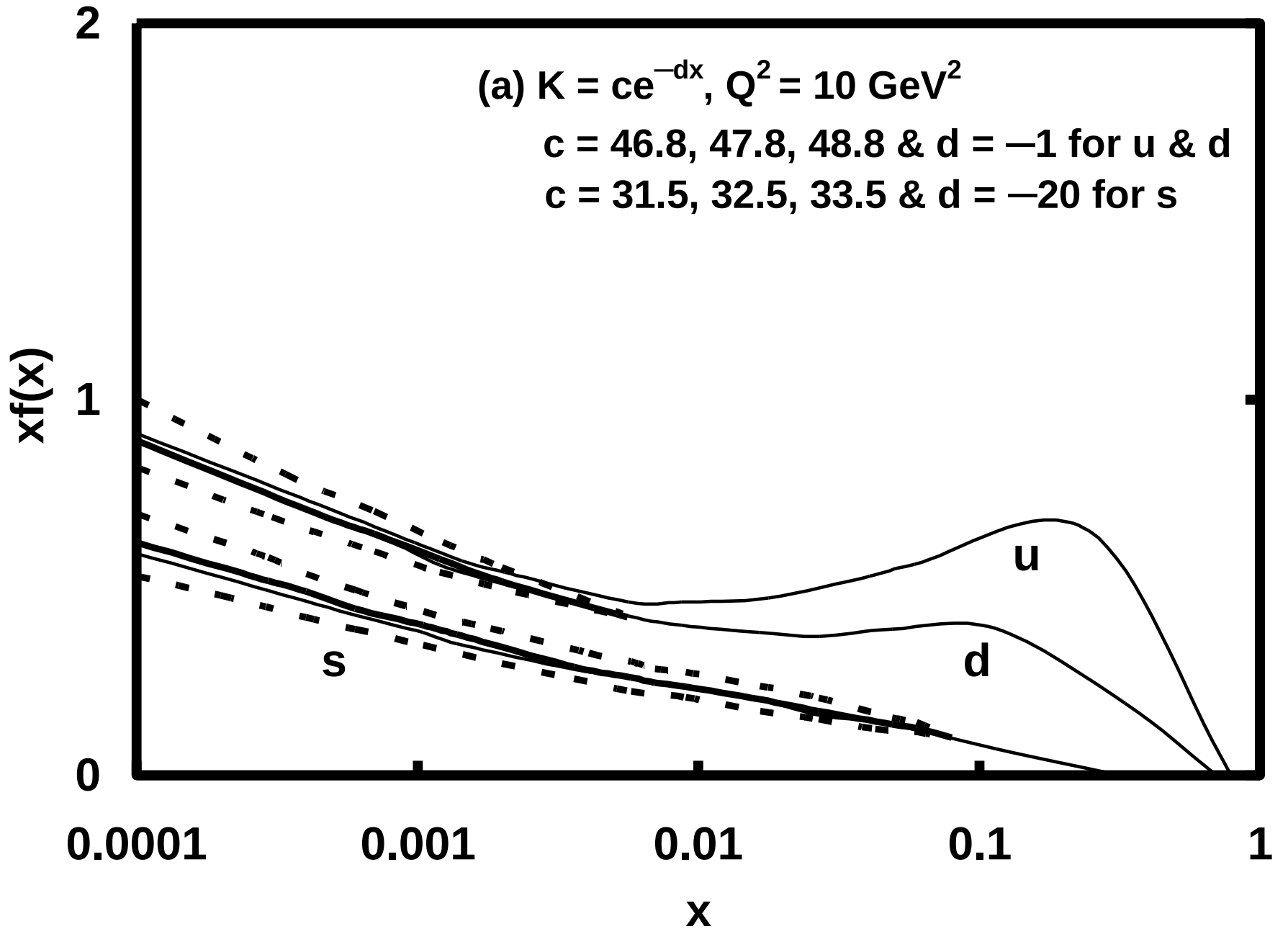

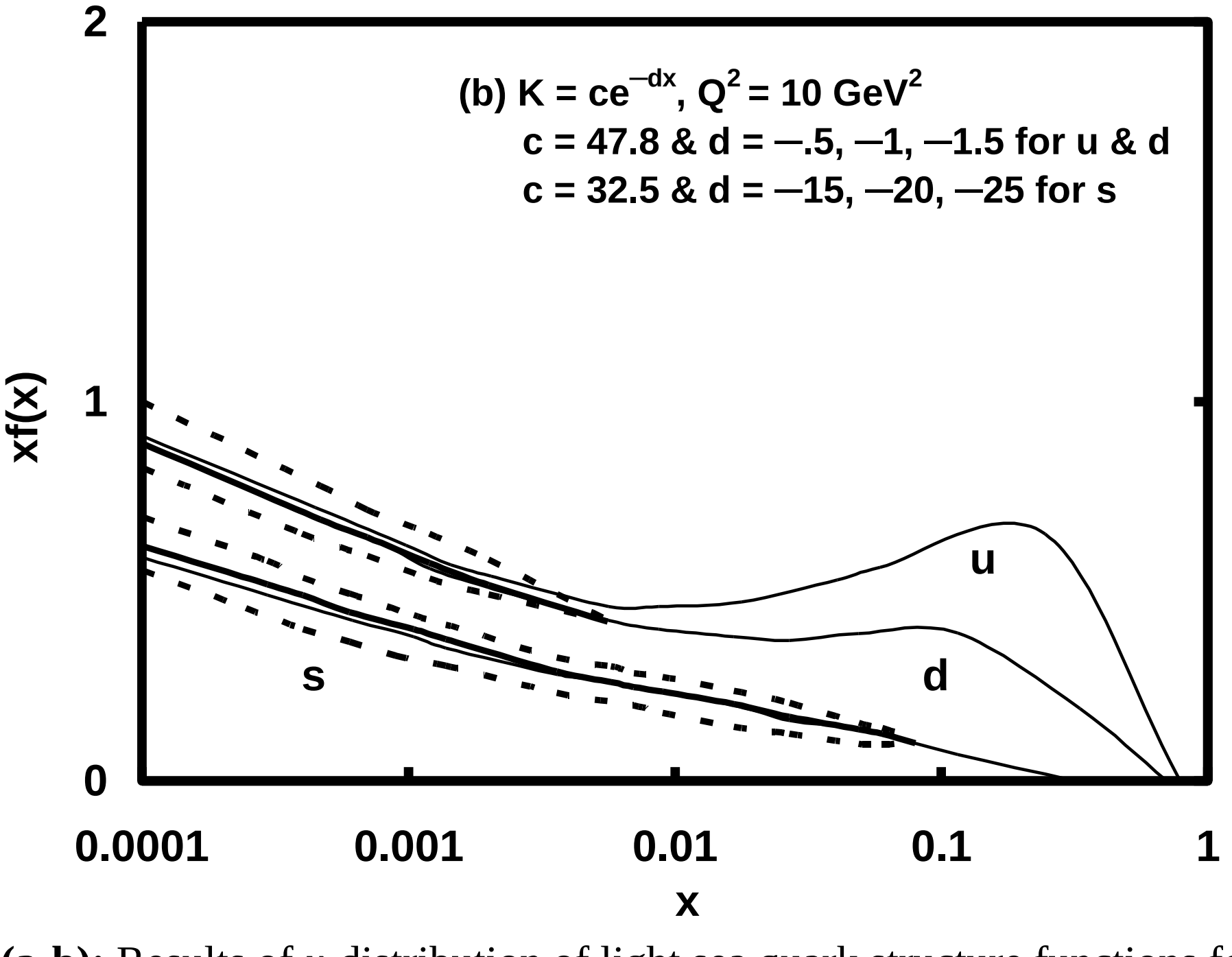

**Fig. 5(a-b):** Results of *x*-distribution of light sea quark structure functions for $K(x) = ce^{-dx}$, where '*c*' and '*d*' are constants for $Q^2 = 10$ GeV$^2$, and compare them with recent global parameterizations (thin solid lines) [8] in the relation $\beta=\alpha^y$ for *y* minimum (thick solid lines). In the same figures we present the sensitivity of our results (dashed lines) for different values of '*c*' and '*d*'.

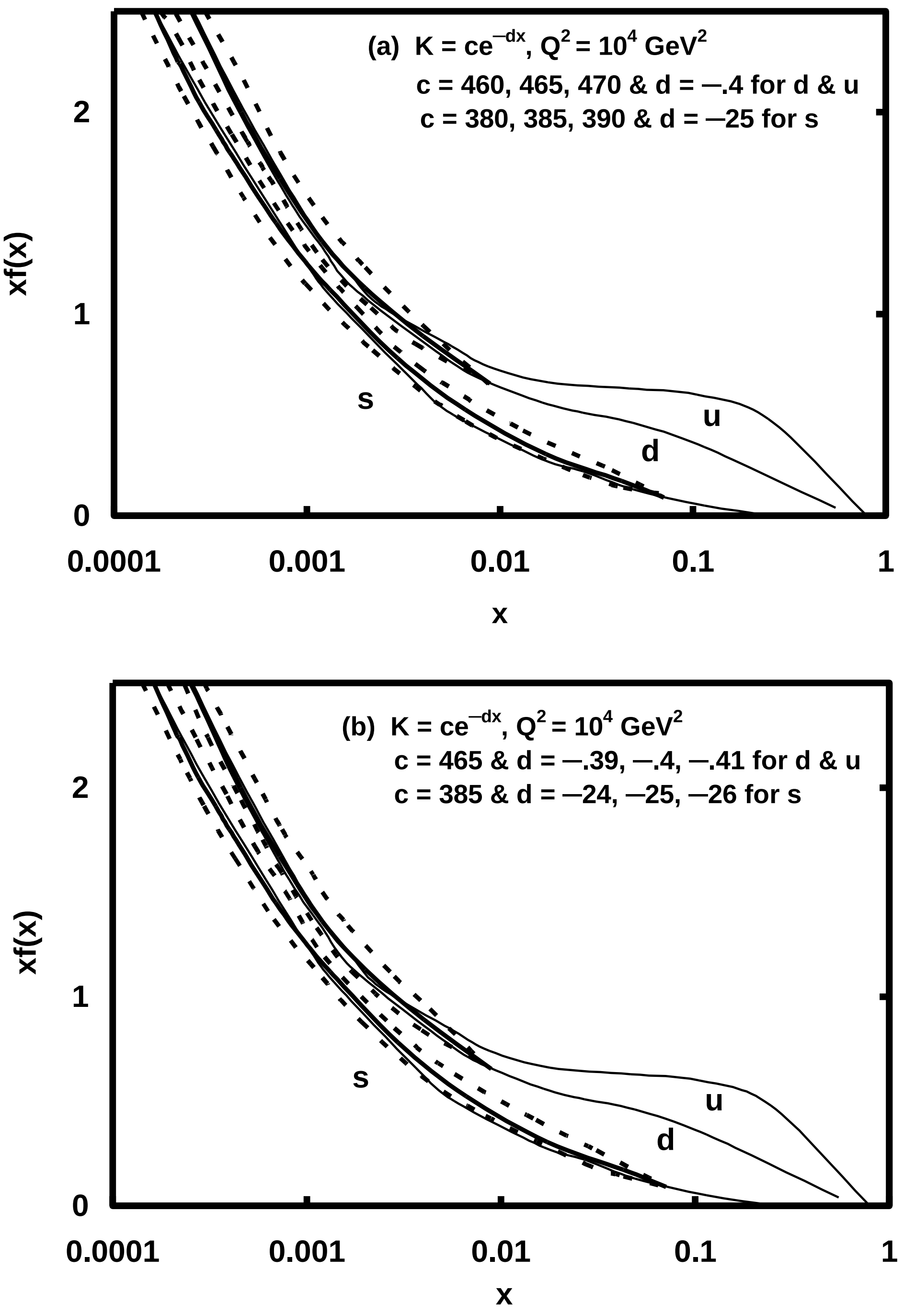

**Fig. 6(a-b):** Results of *x*-distribution of light sea quark structure functions for $K(x)$ = ce$^{-dx}$, where '*c*' and'*d*' are constants for $Q^2 = 10^4$ GeV$^2$, and compare them with recent global parameterizations (thin solid lines) [8] in the relation $\beta=\alpha^y$ for *y* minimum (thick solid lines). In the same figures we present the sensitivity of our results (dashed lines) for different values of '*c*' and '*d*'.

In figure 5(b) and figure 6 (b), we present the sensitivity of our results (dashed lines) for different values of'd' at fixed value of '*c*'. Here we take *c* = 47.8, 32.5 in figure 5(b) and *c* = 465, 385 in figure 4(b). We observe that at *d* = -1, & -20, agreement of the results (thick solid lines) with parameterizations data is found to be good in figure 5(b) and at *d* = -.4 & -25, agreement of the results (thick solid lines) with parameterizations data is found to be excellent in figure 6(b). If value of'd' is increased or decreased the curve goes upward or downward direction. But the nature of the curve is similar.

We observed that for *x*- evolutions of light sea quark structure functions, results for *y* minimum and maximum in $\beta = \alpha^y$ relation in LO have not any significance difference.

It is to be noted that unique solutions evolution equations for valence and light sea quark structure functions are same with particular solutions for *y* maximum in $\beta = \alpha^y$ relation in LO.

From our above discussion, it has been observed that though we can derive a complete unique *t*-evolution for valence and light sea quark structure functions in LO, yet we can not establish a complete unique *x*-evolution for light sea quark function in LO. $K(x)$ is the relation between light sea quark and gluon structure functions may be in the forms of a constant, an exponential function of *x* or a power in *x* and they can equally produce required *x*-distribution of light sea quark. But unlike many parameter arbitrary input *x*-distribution functions generally used in the literature, our method required only one or two such parameter. On the other hand, the explicit form of $K(x)$ can actually be obtained only by solving coupled DGLAP evolution equations for singlet and gluon structure functions, and works are going on in this regard.